\documentclass[aps,prl,twocolumn,showpacs]{revtex4}

\newcommand{\nc}{\newcommand}
\nc{\be}{\begin{equation}} \nc{\ee}{\end{equation}}
\nc{\bea}{\begin{eqnarray}} \nc{\eea}{\end{eqnarray}}
\nc{\bi}[1]{\bibitem{#1}}
\begin{document}


\leftline{\hspace{5.4in} UFIFT-QG-08-01}

\vskip 0.2in

\title{A Decisive test to confirm or rule out the existence of dark
matter emulators using gravitational wave observations}

\author{E. O. Kahya }
\email[]{emre@phys.ufl.edu} \affiliation{Department of Physics,
University of Florida,
             Gainesville, FL 32611, USA}

\date{\today}

\begin{abstract}

We consider stable modified theories of gravity that reproduce
galactic rotation curves and the observed amount of weak lensing
without dark matter. In any such model gravity waves follow a
different geodesic from that of other massless particles. For a
specific class of models which we call ``dark matter emulators,''
over cosmological distances this results in an easily detectable
and difference between the arrival times of the pulse of gravity
waves from some cosmic event and those of photons or neutrinos.
For a repeat of SN 1987a (which took place in the Large Magellanic
Cloud) the time lag is in the range of days. For the recent gamma
ray burst, GRB 070201 (which seems to have taken place on the edge
of the Andromeda galaxy) the time lag would be in the range of
about two years.

\end{abstract}

\pacs{98.80.Cq, 98.80.Hw, 04.62.+v}

\maketitle

%
%


\textbf{1. Introduction}

 Zwicky\cite{Zwi} first recognized the missing mass problem in 1933. Thirty
years later the inconsistency between the observed rotation
velocities in the disk of spiral galaxies, and the predictions of
Newtonian dynamics was demonstrated by Rubin, Ford and
Thonnard\cite{RTF,RFT1,RFT2}. And recently weak lensing showed us
the need for extra
matter\cite{TVW,FKSW,SKFBW,CLKHG,Mellier,WTKAB}.

 Dark matter has been the usual explanation for the failure of
the Newtonian description of the observed phenomena. It assumes
general relativity and posits the existence of vast distributions
of unseen matter to explain the observed cosmic motions and
lensing. There are plenty of candidates for dark matter \cite{DM}:
axions, WIMPs, MACHOs, sterile neutrinos, etc. But despite all the
observational effort we still have not directly detected dark
matter.

 Another solution is modifying general relativity
on cosmological scales. In 1983, Milgrom\cite{Milg} proposed
Modified Newtonian Dynamics (MOND) which was designed to explain
the Tully-Fisher relation between the luminosity of a spiral
galaxy and the peak velocity in its rotation curve\cite{TF}. In
its original form MOND was not a full metric theory and so could
not answer questions about gravitational lensing and cosmology.
However, Bekenstein has recently proposed a relativistic extension
of MOND called TeVeS (for ``Tensor-Vector-Scalar'') \cite{Bek}.
TeVeS gives a plausible amount of weak lensing and, when its
predictions for cosmological evolution are compared with data the
agreement is better than many thought possible
\cite{Skordis,AFZ,Dodelson,Bourliot,BEF}.

TeVeS is an impressive achievement, however, it contains a number
of free parameters. Nor is TeVeS the only modification of gravity
with a claim to explain cosmic motions and lensing without dark
matter\cite{JWM,BrM,MT}. This introduces a large amount of
model-dependence into the task of comparing general relativity +
dark matter with modified gravity. The possibility for a
model-independent test is suggested by the fact that both
TeVeS\cite{Bek} and its rival SVTG\cite{JWM}
(``Scalar-Vector-Tensor-Gravity'') have different metrics'
coupling to ordinary matter and to gravity waves in the weak field
limit. The metric coupled to matter is the one general relativity
would predict if dark matter had been present; the metric coupled
to weak gravity waves is the one general relativity gives without
dark matter. This feature is not an accident. It results from the
tension between recovering solar system tests and a powerful no-go
theorem about single-metric modifications of
gravity\cite{SW1,SW2}. We will shortly review this theorem but let
me proceed by defining as a {\it dark matter emulator} any
modified gravity theory with the bi-metric feature just described.

 We propose a decisive test using gravitational wave observations
of any cosmic event for which an optical or neutrino signal is
also detected. Because the pulse of gravity waves from the event
will follow a different geodesic from the pulse of photons or
neutrinos, there will be a very significant time lag. The time lag
for SN1987A is of the order of days and for GRB070201 is of the
order of months. These lags are typical for events in the general
vicinity of the Milky Way and Andromeda, respectively. So the
detection of a single simultaneous gravity wave and either photon
or neutrino pulse would refute the entire class of dark matter
emulators. Conversely, the detection of the predicted time lag
would refute the hypothesis of dark matter.

 There have been various proposed tests of alternative theories
of gravity using gravitational wave observations. Most of these
tests are in the strong field regime(See Ref \cite{AFY} and
references therein for tests of strong field gravity and
Ref \cite{DKM,KHM} for other tests of gravity). This is the first
proposed test using gravitational wave observations to test any
modifications to gravity in the dark matter dominated regime,
and which can be directly tested using data from ground based
interferometers such as LIGO, VIRGO etc.

\vskip 0.3in

\textbf{2. Soussa-Woodard No-Go Theorem}

 In this section, we will just state the assumptions and the consequences of
Soussa-Woodard No-Go Theorem. The assumptions are as follows:

\begin{itemize}
\item{Gravitational force is carried by the metric, and the source
is the usual energy momentum tensor, \be {\cal E}_{\mu \nu} =
\frac{8\pi G}{c^4} T_{\mu\nu} \; . \label{Ein} \ee The left hand
side of the equation is not necessarily Einstein's tensor. It can
involve higher derivatives, or even more complicated nonlocal
functionals of the metric.} \item{The theory of gravitation is
generally covariant.} \item{MOND force is realized in the weak
field perturbation theory.} \item{The theory of gravitation is
absolutely stable.} \item{Electromagnetism couples conformally to
gravity.} \end{itemize}

The theorem is that these five assumptions are inconsistent with
the observed amount of weak gravitational lensing. Therefore, any
modified gravity theory which attempts to supersede dark matter
must violate one or more of the five assumptions. Dark matter
emulators violate the first assumption by having the MOND force
mediated by some other field in the metric which couples to
ordinary matter. On the other hand, recovering solar system tests
predisposes the action of any such model to possess an
Einstein-Hilbert term, and it is this which controls the
propagation of tensor gravity waves in the weak field regime.

\vskip 0.3in

\textbf{3. A Decisive Test of Dark Matter Emulators}

The idea is very simple: Let us take a cosmic event that produces
simultaneous pulses of gravitational waves and photons/neutrinos.
If general relativity + dark matter is correct then all massless
particles follow the same geodesics and the pulses should arrive
at the same time. If physics is instead described by a dark matter
emulator, then gravity waves will follow a different geodesic from
photons and neutrinos. Both geodesics can be worked out
generically, assuming only that the dark matter emulator
reproduces observed phenomena without dark matter. The difference
is small, but it builds up to an enormous time lag for propagation
over cosmic distances.

It is simple to see that the pulse of gravity waves must arrive
before the photon and neutrino pulses. Gravitational waves couple
to the metric which is predicted by general relativity using only
the visible matter. Ordinary matter couples to a metric that may
be a combination of the metric just alluded to and additional
vector and/or tensor fields, but whose net effect is that of a
normal metric with visible visible matter and the vastly larger
halo of dark matter. Hence the photon and neutrino pulses must
experience more Shapiro delay\cite{IS} than the pulse of gravity
waves \cite{MJL,KT}.

The first calculation of this kind was done for Supernova
1987A\cite{KW1}. It was found that the time lag is of the order of
days which would have been easy to distinguish from the optical
and neutrino pulses, if sufficiently sensitive gravity wave
detectors were operating at that time. A minor point but one of
great significance to future observations is that the actual time
lag reported in the earlir work\cite{KW1} was negative. That is,
we predicted the gravity wave pulse to arrive {\it after} the
optical and neutrino pulses, rather than the correct result of
before. Had the predicted speed of gravity been less than that of
light the dark matter emulator would have already been ruled out by the ingenious
argument of Moore and Nelson\cite{MN}. The problem arose from an
incorrect assumption about the parameter $r_0$ which gives the
distance at which the dark matter density vanishes in the
isothermal halo model. Considering the Milky Way in company with
other galaxies makes it clear that this parameter must be hundreds
of kiloparsecs rather than the value of 8 kpc which we had assumed
\cite{DKW}.

And a similar calculation is being done for GRB070201\cite{DKW}.
This short duration Gamma Ray Burst (GRB) is thought to have
happened on the edge of the Andromeda galaxy. It has been
suggested that Short Gamma Ray Bursts arise from compact binary
mergers, in which case a detectable pulse of gravity waves would
have been produced \cite{EN,LIGO}. Initial estimates, based on the
isothermal halo model for dark matter profile, show that the time
lag is at the order of two years. Taking account of the reported
uncertainty in angular position (assuming an origin at the
distance of Andromeda) results in variations of about a week in
the predicted time lag within the context of the isothermal halo
model. The calculation for different dark matter profiles is being
done to see how much it depends on the particular model.

The first reason for doing the calculation for GRB 070201 is its
having the crucial property of simultaneous pulses of
gravitational waves and photons. The second reason is its source
being at the edge of Andromeda galaxy, which is not too far to
preclude a detectable gravity wave signal if the event was in fact
a compact binary merger. But the most important reason is that
searches were conducted with the LIGO detectors for gravitational
wave signals within a 180 second window around the GRB and no
gravitational waves were detected. This implies that compact
binary mergers up to 3 Mpc are excluded or this event(if located
in M31) is a SGR\cite{LIGO,LIGO2}.

Of course GRB 070201 may not have derived from a compact binary
merger, or it may have been further than 3 Mpc. However, it might
also be that a dark matter emulator is the correct theory of
gravity and that the reason the gravity wave pulse was not
detected is that it arrived two years before the time interval
which was searched. If that is the case, one should look at the
old data, where the amount of time lag can be calculated with a
reasonable error. The gravitational wave signal and the antenna
patterns would be same as in GR, with no extra polarization
states. The time delay of the gravitational waves between Hanford
and Livingston LIGO detectors would be same as the light travel
time. Hence no modifications to the usual data analysis
techniques\cite{LIGO} are needed to do this search in these dark
matter emulator models.

There are many candidate sources for similar tests of modified
gravity models, such as close GRBs, SGRs, supernova explosions,
pulsar glitches, optical transients\cite{CWS}, etc. The time delay
between production of gravitational waves and electromagnetic
counterparts in almost all proposed models is much smaller than an
hour\cite{JS}. Due to the uncertainty in measuring the distance of
to the cosmic events, one would prefer closer objects to get more
precise estimates for the time lag. A future project would be to
produce a rough graph of the time lag as a function of position in
the Milky Way-Andromeda neighborhood which could serve as a guide
for observers as to time periods to be searched for the gravity
wave pulses associated with optical and/or neutrino pulses. The
detection of a gravity wave pulse arriving at the predicted time
before its optical or neutrino counterpart would be strong
evidence against dark matter. Conversely, the observation of
simultaneous arrival times would refute the entire class of dark
matter emulators.

\vskip 0.3in

\textbf{4. Conclusion}

Although the usual explanation for cosmic motions and the observed
amount of weak lensing is vast distributions of unseen matter, it
is also possible that general relativity is not the correct theory
of gravity on cosmic scales. A powerful no-go theorem restricts
any modified gravity theory that attempts to supplant dark matter.
A certain class of models called {\it dark matter emulators} evade
this theorem by possessing two metrics. The first is numerically
equal to the metric that general relativity would produce with
dark matter; this is the metric which couples to ordinary matter.
The second metric is numerically equal to what general relativity
would give without dark matter; this is the metric which couples
to weak gravitational waves.

Because gravity waves follow different geodesics from those of
other massless particles, there will be a substantial time lag
between the arrival of simultaneous pulses of photons or neutrinos
and gravity waves. The gravity wave pulse will always arrive
first. Explicit computations have been done for SN 1987A and for
GRB 070201, giving time lags of several days and about two years,
respectively.

In all triggered gravitational wave searches  in LIGO/VIRGO so far
it has been assumed that the gravity wave pulse was coincident
with the corresponding pulses of photons and/or neutrinos
\cite{ZM}. If dark matter does not exist and a dark matter
emulator is right, this assumption completely breaks down and one
should also look for coincidence at non-zero times.

The gravity wave detectors won't give much directional inofrmation,
although the waveform will tell a little about the type of the
source It would only be after seeing many cases of seeing gravity
wave signals, followed by a plausible source that one would
proclaim verification of a {\it dark matter emulator}. On the
other hand, a single coincident signal would rule out
{\it dark matter emulators}. However, even there it would not
confirm dark matter, it would just mean that, if modified gravity
explains dark matter, it must be through some other way.

To conclude, if this effect is real, it might mean that general
relativity is not the right theory of gravity and might also mean
that there is no dark matter, which would change our understanding
of cosmology. A motivation of doing an experiment is testing
viable theories and we think that this a very important and a
doable test.

\vskip 0.3in

\textbf{Acknowledgements}

I am grateful to Richard Woodard and Shantanu Desai
for stimulating discussions and encouraging me for
this work. I thank John Beacom, Marie-Anne Bizouard,
Jim Fry, Patrice Hello, Bence Kocsis, Szabi Marka,
Zsusza Marka, Dave Seckel, Pierre Sikivie, Chris Stubbs,
Bernard Whiting for illuminating comments and discussions.
I also am grateful to the organizers of the 12th
Gravitational Wave Data Analysis Workshop for giving
me a chance for useful discussions. This work was partially
supported by NSF grants PHY-244714 and PHY-
065305 and by the Institute for Fundamental Theory at
the University of Florida.

\end{document}